\documentclass[12pt]{article}%
\usepackage{amsfonts}
\usepackage{amsmath}
\usepackage{amssymb}
\usepackage{graphicx}%
\newtheorem{theorem}{Theorem}

\newtheorem{corollary}[theorem]{Corollary}

\newtheorem{definition}[theorem]{Definition}

\newenvironment{proof}[1][Proof]{\noindent\textbf{#1.} }{\ \rule{0.5em}{0.5em}}
\begin{document}

\title{Combinatorial Laplacians and Positivity Under Partial Transpose}
\author{Roland Hildebrand\thanks{roland.hildebrand@imag.fr}\\LJK, Universit\'{e} Joseph Fourier, Tour IRMA, \\51 rue des Math\'{e}matiques, 38400 St. Martin d'H\`{e}res, France
\and Stefano Mancini\thanks{stefano.mancini@unicam.it}\\Dipartimento di Fisica, Universit\`{a} di Camerino, \\Via Madonna delle Carceri, 9, I-62032 Camerino, Italy
\and Simone Severini\thanks{ss54@york.ac.uk}\\Department of Mathematics and Department of Computer Science, \\University of York, Y010 5DD York, United Kingdom}
\maketitle

\begin{abstract}
Density matrices of graphs are combinatorial laplacians normalized to have
trace one (Braunstein \emph{et al.} \emph{Phys. Rev. A,} \textbf{73}:1, 012320
(2006)). If the vertices of a graph are arranged as an array, then its density
matrix carries a block structure with respect to which properties such as separability
can be considered. We prove that the so-called degree-criterion, which was conjectured to be
necessary and sufficient for separability of density matrices of graphs, is
equivalent to the PPT-criterion. As such it is not sufficient for testing the separability
of density matrices of graphs (we provide an explicit example).
Nonetheless, we prove the sufficiency when one of the array dimensions has
length two (for an alternative proof see Wu, \emph{Phys. Lett. A}\textbf{351} (2006), no. 1-2,
18--22).

Finally we derive a rational upper bound on the concurrence of density
matrices of graphs and show that this bound is exact for graphs on four vertices.
\end{abstract}

\section{Introduction}

Introduced by Braunstein \emph{et al.} \cite{braunstein06a,
braunstein06b}, density matrices of graphs are simply combinatorial
laplacians normalized to have unit trace (the normalization consists
of dividing the nonzero entries by twice the cardinality of the edge
set). In this way, to any graph $G$ (with labeled vertices) 
can be associated a
specific mixed quantum state (identified with its matrix
representation), which is then called the \emph{density matrix of
}$G$. If the vertices of a graph are arranged in a multi-dimensional
array, then the density matrix of the graph carries a block
structure, which can be associated with a split of the quantum
system into subsystems. Each array dimension will then correspond to
one subsystem, and the length of the array dimension will equal the
number of pure states the subsystem can assume. It is useful to
remark that the combinatorial properties of the graph $G$ up to
isomorphism do not always characterize its density matrix and
therefore do not specify the physical properties of the state. This
explains why we need to consider labeled graphs. In other words, we
assume that graphs with different adjacency matrices (even if
belonging to the same isomorphism class) have different density
matrices and then correspond to different quantum states, whose
properties can be radically different.

Studying density matrices of graphs with the tool-box provided by quantum
mechanics has a twofold role: from the perspective of combinatorics, this
interface can be fruitful in uncovering and re-defining graph theoretic
properties; from the perspective of quantum mechanics, density matrices of
graphs can be seen as \textquotedblleft simple\textquotedblright\ and
\textquotedblleft highly symmetric\textquotedblright\ states. Observed under
this light, density matrices of graphs provide a restricted testing ground for
better understanding techniques and concepts employed in more general
settings. Such an approach has particular value, when considering the
particular kind of developments in quantum physics and its applications that
we are experiencing today.

Indeed, the study of finite-dimensional states is important in quantum
information processing. This is the multidisciplinary area whose goal is to
understand and exploit the information encoded in quantum states (see Nielsen
and Chuang \cite{nielsen00} for a monograph on the subject; see Alber
\emph{et al.} \cite{alber01} for a collection of overviews). The basis of
this field consists of interpreting certain quantum physical entities as
information carriers and their evolution in time as information processing
dynamics. Such a view is giving rise to a number of discoveries and successful
real world applications, the most popularized examples being quantum
communication and quantum computing. The main ingredient which most likely is
responsible for the \textquotedblleft quantumness\textquotedblright\ is the
concept of \emph{entanglement}, a property associated to certain quantum states.

While at the early stages of quantum physics entanglement was considered a
mystery, nowadays it is recognized as a precious resource, difficult to create
and to preserve. Defining entanglement is not an easy task
(Bru\ss \ \cite{brus02} compiles an eloquent list of definitions). Speaking
about entanglement makes sense only if one considers a system composed of at
least two subsystems. The rough idea is that if the two parties (or,
equivalently, subsystems) are \emph{entangled} then a complete description of
the whole system does not imply a complete description of the parts and
\emph{vice versa}. So, two entangled systems present some kind of
\emph{correlation} that does not appear to occur in the realm of classical
mechanics, where complete information on the system implies a complete
description of its individual parts.

>From the mathematical point of view, the theory of entanglement is rich and diversified. It has branches in
geometry, knots, Lie groups, positive maps, combinatorics, convex optimization, \emph{etc}. The main problems
are: \emph{(i)} determine whether a given quantum state is entangled; \emph{(ii) }determine how much
entanglement is in a given quantum state; \emph{(iii) }determine the \textquotedblleft
quality\textquotedblright\ of entanglement (\emph{e.g.}, the problem of distillability).

As we mentioned above, density matrices of graphs are a restricted set in
which these tasks can have a special treatment. Specifically, Braunstein
\emph{et al.} \cite{braunstein06a, braunstein06b} and Wu \cite{wu06}
considered the \emph{Quantum Separability Problem (QSP) }for these matrices.
The QSP is the computational problem of deciding whether a given quantum state
is entangled or not, that is \emph{separable }(Ioannu 2006 \cite{ioannou06} is
a recent review). The QSP is equivalent to an instance of a combinatorial
optimization problem called the Weak Membership Problem and defined in
Gr\"{o}tschel \emph{et al. }\cite{gro88}. In its complete generality the QSP
is NP-hard (Gurvits \cite{gurvits03}).

There is some evidence that the QSP for density matrices of graphs might be
easier than for general density matrices. A simple necessary condition for
separability is that the degrees of the vertices of $G$ are the same as the degrees
of the vertices of another graph, $G^{\prime}$,
obtained from $G$ by mean of a simple operation acting on the edges. The
operation is a combinatorial analogue of the linear algebraic partial
transposition. In fact, the graph $G^{\prime}$ will be called here
\emph{partially transposed graph}. The condition for separability will be
called here \emph{degree-criterion}. Since the partial transposition is
centrally involved in the famous \emph{Peres-Horodeckis' criterion} for
separability of general states \cite{peres96, ho96} (also called \emph{Positivity under
Partial Transpose Criterion}, or, for short, \emph{PPT-criterion)}, it is
natural to investigate the relation between the degree-criterion and the
PPT-criterion when applied to density matrices of graphs.

In this paper, we give an elementary proof that the two criteria are
equivalent for density matrices of graphs. We also exhibit a simple example
showing that the degree-criterion is not sufficient for testing separability
of density matrices of disconnected graphs (that is, graphs with more than one
connected component). Additionally, we verify the sufficiency of the degree-criterion
when the dimension of one of the parties is two, therefore giving an alternative proof
of a result of Wu \cite{wu06}. There are four sections in the paper. The above
observations are exposed in Section 2, after providing the necessary notions
and terminology. Section 3 is devoted to point \emph{(ii)} above. In
particular, we focus on the concurrence. This is a quantity associated to
every density matrix and it is strictly larger than zero for entangled states
(Hill and Wootters \cite{hill97}). We derive a simple upper bound on the
concurrence of density matrices of graphs and show the exactness of this bound
for graphs with four vertices. In Section 4 we draw some conclusions.

\section{The degree-criterion and the PPT-criterion for density matrices of
graphs}

The purpose of this section is to shed further light onto the QSP of density
matrices of graphs. First, we state formally the QSP and define the
PPT-criterion. We then recall the notion of combinatorial laplacian. We define
the degree-criterion and we prove its equivalence with the PPT-criterion. We
conclude the section by showing that the degree-criterion is necessary and
sufficient for testing separability of density matrices of graphs in
$\mathbb{C}_{A}^{2}\otimes\mathbb{C}_{B}^{q}$ (see also Wu \cite{wu06}). Our
reference on graph theory is Godsil and Royle \cite{godsil01}.

\subsection{The quantum separability problem}

In the axiomatic formulation of quantum mechanics in Hilbert space, the state
of a quantum mechanical system, associated to the $n$-dimensional Hilbert
space $\mathcal{H}\cong\mathbb{C}^{n}$, is identified with an $n\times n$
positive semidefinite, trace-one hermitian matrix, called a \emph{density
matrix}. In Dirac notation, a unit vector in a Hilbert space $\mathcal{H}%
\cong\mathbb{C}^{n}$ is denoted by $\left\vert \psi\right\rangle $, where
$\psi$ is simply a label; given the vectors $\left\vert \varphi\right\rangle
,\left\vert \psi\right\rangle \in\mathcal{H}$, the linear functional sending
$\left\vert \psi\right\rangle $ to the inner product $\left\langle
\varphi|\psi\right\rangle $ is denoted by $\left\langle \varphi\right\vert $.
(We could easily avoid the Dirac notation here, but we use it to be coherent
with the standard literature.) For any unit vector $\left\vert \psi
\right\rangle \in\mathcal{H}$, the projector on $\left\vert \psi\right\rangle
$ is the hermitian matrix $P\left[ \left\vert \psi\right\rangle \right]
:=|\psi\rangle\langle\psi|$, which is called \emph{pure state}. Every density
matrix can be written as a weighted sum of pure states (with real nonnegative
weights summing up to 1); if the sum has more than one component then the
state is said to be \emph{mixed}. By this definition, the decomposition of a
mixed density matrix into pure states is not necessarily unique. A matrix of
the form $P\left[ \left\vert \psi\right\rangle \right] \otimes P\left[
\left\vert \varphi\right\rangle \right] $ is called a \emph{product state},
where the symbol \textquotedblleft$\otimes$\textquotedblright\ denotes the
Kronecker or tensor product. Let $\emph{S}_{A}$ and $\emph{S}_{B}$ be two quantum
mechanical systems associated to the $p$-dimensional and $q$-dimensional
Hilbert spaces $\mathcal{H}_{A}\cong\mathbb{C}_{A}^{p}$ and $\mathcal{H}%
_{B}\cong\mathbb{C}_{B}^{q}$, respectively. The composite system
$\emph{S}_{AB}$, which consists of the subsystems $\emph{S}_{A}$ and
$\emph{S}_{B}$, is associated to the Hilbert space $\mathcal{H}_{AB}%
\cong\mathbb{C}_{A}^{p}\otimes\mathbb{C}_{B}^{q}$. The density matrix $\rho$
of $\emph{S}_{AB}$ is said to be \emph{separable} if%
\[
\rho=\sum_{i=1}^{N}p_{i}P\left[ \left\vert \psi_i\right\rangle _{A}\right]
\otimes P\left[ \left\vert \varphi_i\right\rangle _{B}\right] ,
\]%
\[
\text{where }p_{i}\geq0\text{, for every }i=1,2,...,N,\text{ and }\sum
_{i=1}^{n}p_{i}=1;
\]
the projectors $P\left[ \left\vert \psi_i\right\rangle _{A}\right] $ and
$P\left[ \left\vert \varphi_i\right\rangle _{B}\right] $ are product states
acting on $\mathcal{H}_{A}$ and $\mathcal{H}_{B}$, respectively. A density
matrix $\rho$ is said to be \emph{entangled} if it is not separable. Entangled
states cannot be prepared from separable states by mean of operations acting
locally on the subsystems. Although the definition given here refers to
exactly two parties, entanglement can be defined as well for systems composed
of many subsystems.

\subsection{The PPT-criterion}

The PPT-criterion is based on the notion of partial transpose. This is a
common and important notion in the study of entanglement. Let $\rho$ be a
density matrix acting on the Hilbert space $\mathcal{H}_{AB}\cong\mathbb{C}%
_{A}^{p}\otimes\mathbb{C}_{B}^{q}$. Let
\[%
\begin{tabular}
[c]{lll}%
$\left\{ \left\vert u_{1}\right\rangle ,\left\vert u_{2}\right\rangle
,...,\left\vert u_{p}\right\rangle \right\} $ & and & $\left\{ \left\vert
w_{1}\right\rangle ,\left\vert w_{2}\right\rangle ,...,\left\vert
w_{q}\right\rangle \right\} $%
\end{tabular}
\ \ \ \
\]
be orthonormal bases of $\mathbb{C}_{A}^{p}$ and $\mathbb{C}_{B}^{q}$,
respectively. Let $\{\left\vert v_{1}\right\rangle ,\left\vert v_{2}%
\right\rangle ,...,\left\vert v_{n}\right\rangle \}$ be an orthonormal basis
of $\mathcal{H}_{AB}$, where $n = pq$. Alternatively, we index these basis vectors with pairs $(k,l)$.
These vectors are taken as follows:
\[ \left\vert v_{(k-1)q+l}\right\rangle = \left\vert v_{k,l}\right\rangle = \left\vert u_{k}\right\rangle \left\vert
w_{l}\right\rangle, \quad k = 1,\dots,p;\ l = 1,\dots,q.
\]
The \emph{partial transpose} of $\rho$ with respect to the system
$\emph{S}_{B}$ is the $pq\times pq$ matrix, denoted by
$\rho^{\Gamma_{B}}$, with the $\left(
i,j;i^{\prime},j^{\prime}\right) $-th entry defined as follows:
\[
\lbrack\rho^{\Gamma_{B}}]_{i,j;i^{\prime},j^{\prime}}=\langle u_{i}|\langle
w_{j^{\prime}}|\rho|w_{j}\rangle|u_{i^{\prime}}\rangle,
\]
where $1\leq i,i^{\prime}\leq p$ and $1\leq j,j^{\prime}\leq q$. The density
matrix of $\emph{S}_{AB}$ can be written as
\begin{equation}
\rho=\left(
\begin{array}
[c]{ccc}%
A_{11} & \ldots & A_{1p}\\
\vdots & \ddots & \vdots\\
A_{p1} & \ldots & A_{pp}%
\end{array}
\right) , \label{1}%
\end{equation}
with $q\times q$ matrices $A_{ij}$ acting on the space $\mathbb{C}_{B}^{q}$.
The partial transpose is then realized by transposing all these matrices:%
\[
\rho^{\Gamma_{B}}=\left(
\begin{array}
[c]{ccc}%
A_{11}^{T} & \ldots & A_{1p}^{T}\\
\vdots & \ddots & \vdots\\
A_{p1}^{T} & \ldots & A_{pp}^{T}%
\end{array}
\right) .
\]
If $\rho$ is separable then $\rho^{\Gamma_{B}}\geq0$ (see Peres \cite{peres96}%
). However, the converse is not necessarily true, since there exist
entangled states with positive partial transpose (the so-called
\emph{bound entangled states}). The failure of the PPT-criterion is
then the failure of an operational characterization of entangled
states, which is computationally simple verify. The PPT-criterion is
necessary and
sufficient for separability of density matrices acting on $\mathbb{C}_{A}%
^{2}\otimes\mathbb{C}_{B}^{2}$ or $\mathbb{C}_{A}^{2}\otimes\mathbb{C}_{B}%
^{3}$ (Horodecki \emph{et al.} \cite{ho96}); it is also necessary and
sufficient for certain infinite-dimensional states (Simon \cite{simon00}; Duan
\emph{et al. }\cite{duan}; Mancini and Severini \cite{mancini06} is a brief
review). It is important to mention that only one other (operational) criterion
is known for detecting entanglement:\ the \emph{realignment criterion}
(Rudolph \cite{rudolph02}; Chen \cite{chen02}). It can detect bound
entanglement, but for some states it is weaker than the PPT-criterion.
Unfortunately, it can be checked numerically that the two criteria together do not
solve the QSP for all states (see Horodecki and Lewenstein \cite{ho00}).
Generally, the operational characterization of entanglement is an open problem.

\subsection{Combinatorial laplacians}

In this subsection we recall the notion of combinatorial laplacian.
A \emph{graph} $G=(V,E)$ is a pair defined in the following way:
$V(G)$ is a non-empty and finite set whose elements are called
\emph{vertices}; $E(G)$ is a non-empty set of unordered pairs of
vertices, which are called \emph{edges}. A \emph{loop} is an edge of
the form $\{v_{i},v_{i}\}$, for some vertex $v_{i}$. We assume that
$E(G)$ does not contain loops. A graph $G$ is said to be \emph{on
}$n$ \emph{vertices} if the number of elements in $V(G)$ is $n$. The
\emph{adjacency matrix} of a graph on $n$ vertices $G$ is an
$n\times n$ matrix, denoted by $M(G)$, having rows and columns
labeled by the vertices of
$G$, and $ij$-th entry defined as follows\footnote{We are considering here only `simple' graphs.}:
\[
\left[ M(G)\right] _{i,j}:=\left\{
\begin{tabular}
[c]{cc}%
$1,$ & if $\{v_{i},v_{j}\}\in E(G);$\\
$0,$ & if $\{v_{i},v_{j}\}\notin E(G).$%
\end{tabular}
\ \ \ \right.
\]
Two vertices $v_{i}$ and $v_{j}$ are said to be \emph{adjacent} if
$\{v_{i},v_{j}\}\in E(G)$. The \emph{degree} of a vertex $v_{i}\in V(G)$,
denoted by $d_{G}(v_{i})$, is the number of edges adjacent to $v_{i}$. The
\emph{degree-sum} of $G$ is defined and denoted by
\[
d_{G}=\sum_{i=1}^{n}d_{G}(v_{i}).
\]
The \emph{degree} \emph{matrix} of $G$ is an $n\times n$ matrix, denoted by
$\Delta(G)$, having $ij$-th entry defined as follows:%
\[
\left[ \Delta(G)\right] _{i,j}:=\left\{
\begin{tabular}
[c]{cc}%
$d_{G}(v_{i}),$ & if $i=j;$\\
$0,$ & if $i\neq j.$%
\end{tabular}
\ \ \ \right.
\]
The \emph{combinatorial laplacian} \emph{matrix }of a graph $G$ (for short,
\emph{laplacian}) is the matrix
\[
L(G):=\Delta(G)-M(G).
\]
According to our definition of graph, $L(G)\neq0$. Moreover, the laplacian of a graph
$G$, scaled by the degree-sum of $G$, has trace one and is semidefinite positive.
As such it has the characteristic features of a quantum mechanical density matrix, hence it would provide a link to quantum states.
On the basis of this observation, we fix the following definition: the
\emph{density matrix of a graph }$G$ is the matrix%
\[
\rho_{G}:=\frac{1}{d_{G}}L(G).
\]
Let $\mathcal{G}_{n}$ be the set of density matrices of graphs on
$n$ vertices. The set $\mathcal{G}_{n}$ is a subset of the set of
all density matrices acting on the $n$-dimensional Hilbert space
$\mathcal{H}_{AB}\cong\mathbb{C}_{A}^{p}\otimes\mathbb{C}_{B}^{q}$,
where $n=pq$. The number of elements in $\mathcal{G}_{n}$ equals the
number of graphs on $n$ vertices, a number that grows
superexponentially in $n$. There are many applications of
laplacians. In particular, their eigensystems are a rich source of
information about graphs (Mohar \cite{mohar88}).

It is important to remark that graphs with different adjacency matrices have different density matrices, even graphs belonging to the same isomorphism class (e.g. those obtained from each other by permutation of the vertices labels). 
In fact, given a graph $G$ with density matrix $\rho_G$, 
if there exist a permutation matrix $P$ such that $P^TM(G)P=M(G')$, then $G\cong G'$. As a consequence $G'$ has density matrix $P^T\rho_G P$. 

Finally, given the density matrix $\rho_G$ of a graph, in order to have a correspondence to a quantum state (density operator), we have to specify the basis in the Hilbert space with respect to which the quantum state (density operator) has $\rho_G$ as matrix representation. This can be done by associating vertices labels to orthonormal vectors.

\subsection{The degree-criterion}

Let $G$ be a graph on $n=pq$ vertices $v_{1},v_{2},\ldots,v_{n}.$
These vertices are represented here as ordered pairs in the following way:
\[
v_{(k-1)p+l}=(u_{k},w_{l})\equiv u_{k}w_{l}, \quad k = 1,\dots,p;\ l = 1,\dots,q.
\]
By respecting this labelling, we associate $G$ to the orthonormal basis
\[
\{|v_{i}\rangle:i=1,2,\ldots,n\}=\{|u_{j}\rangle\otimes|w_{k}\rangle
:j=1,2,\ldots,p;k=1,2,\ldots,q\},
\]
of the Hilbert space $\mathcal{H}_{AB}\cong\mathbb{C}_{A}^{p}\otimes
\mathbb{C}_{B}^{q}$, where
\[%
\begin{tabular}
[c]{lll}%
$\{|u_{j}\rangle:j=1,2,\ldots,p\}$ & and & $\{|w_{k}\rangle:k=1,2,\ldots,q\}.$%
\end{tabular}
\ \ \
\]
are orthonormal bases of the Hilbert spaces $\mathcal{H}_{A}\cong
\mathbb{C}^{p}$ and $\mathcal{H}_{B}\cong\mathbb{C}^{q}$, respectively. The
\emph{partial transpose of a graph} $G=(V,E)$ (with respect to $\mathcal{H}%
_{B}$), denoted by $G^{\Gamma_{B}}=(V,E^{\prime})$, is the graph such that
\[%
\begin{tabular}
[c]{lll}%
$\{u_{i}w_{j},u_{k}w_{l}\}\in E^{\prime}$ & if and only if & $\{u_{i}%
w_{l},u_{k}w_{j}\}\in E$.
\end{tabular}
\ \ \
\]
If $\Delta(G)=\Delta\left( G^{\Gamma_{B}}\right) $ we say that $G$ satisfies
the \emph{degree-criterion}. The following conjecture was proposed in
Braunstein \emph{et al.} \cite{braunstein06b}: a density matrix $\rho_{G}$ of
a graph on $n=pq$ vertices is separable in $\mathbb{C}_{A}^{p}\otimes
\mathbb{C}_{B}^{q}$ if and only if $\Delta(G)=\Delta\left( G^{\Gamma_{B}%
}\right) $. A proof of this conjecture would give a simple method for testing
the separability of density matrices of graphs, as we would only need to check
whether the $n\times n$ diagonal matrices $\Delta(G)$ and $\Delta\left(
G^{\Gamma_{B}}\right) $ are equal. There are \ counterexamples to this
conjecture, when the graph has isolated vertices (that is, vertices not
belonging to any edge). This is the case for the graph $G$ defined
on a $3 \times 3$ array, with laplacian
\[
L(G)=\left(
\begin{array}
[c]{ccc}%
I_{4} & 0 & -I_{4}\\
0 & \vdots & 0\\
-I_{4} & 0 & I_{4}%
\end{array}
\right) ,
\]
where $I_{d}$ is the $d$-dimensional identity matrix. Indeed, $G$ satisfies
the degree-criterion, but $\rho_{G}$ is entangled. For connected graphs, we
don't have any counterexample yet. Next, we list known partial results about separability:

\begin{itemize}
\item Let $\rho_{G}$ be the density matrix of a graph on $n=pq$ vertices. If
$\rho_{G}$ is separable in $\mathbb{C}_{A}^{p}\otimes\mathbb{C}_{B}^{q}$ then
$\Delta(G)=\Delta\left( G^{\Gamma_{B}}\right) $ (Braunstein \emph{et al.
}\cite{braunstein06b}).

\item Let $G$ be a nearest point graph on $n=pq$ vertices. Then the density
matrix $\rho_{G}$ is separable in $\mathbb{C}_{A}^{p}\otimes\mathbb{C}_{B}%
^{q}$ if and only if $\Delta\left( G\right) =\Delta\left( G^{\Gamma_{B}%
}\right) $ (Braunstein \emph{et al. }\cite{braunstein06b}). (It may be worth
recalling the definition of nearest point graph. Consider a rectangular
lattice with $pq$ points arranged in $p$ rows and $q$ columns, such that the
distance between two neighboring points on the same row or in the same column
is $1$. A \emph{nearest point graph} is a graph whose vertices are identified
with the points of the lattice and the edges have length $1$ or $\sqrt{2}$.)

\item Let $G$ and $H$ be two
graphs on $n=pq$ vertices. If $\rho_{G}$ is separable in $\mathbb{C}%
^{p}\otimes\mathbb{C}^{q}$ and $G\cong H$ (that is, $G$ and $H$ are
isomorphic) then $\rho_{H}$ is not necessarily separable in $\mathbb{C}%
^{p}\otimes\mathbb{C}^{q}$. However, there are exceptions, as observed by the
following point (Braustein \emph{et al. }\cite{braunstein06a}).

\item Let $K_{n}$ be the \emph{complete graph} on $n$ vertices. Recall that
the complete graph is the graph with an edge between any pair of vertices. One
can show that, for any $n=pq$, the density matrix $\rho_{K_{n}}$ is separable
in $\mathbb{C}^{p}\otimes\mathbb{C}^{q}$. Notice that for a graph $H$ such
that $M(H)=M(G)\otimes M(G^{\prime})$ for some graphs $G$ and $G^{\prime}$,
the density matrix $\rho_{H}$ is separable. Of course, if a density matrix
$\rho_{G}$ is separable it does not necessarily mean that $M(G)$ is a tensor
product. The \emph{star graph} on $n$ vertices $v_{1},v_{2},...,v_{n}$,
denoted by $K_{1,n-1}$, is the graph whose set of edges is $\{\{v_{1}%
,v_{i}\}:i=2,3,..,n\}$. The density matrix $\rho_{K_{1,n-1}}$ is entangled for
$n=pq\geq4$. So, the separability properties of complete graphs and star
graphs do not depend on the labelling. It is an open problem to determine if
these graphs are the only ones with this property (Braunstein \emph{et al. }
\cite{braunstein06a}).

\item Let $\rho_{G}$ be the density matrix of a graph on $n=2q$ vertices. Then
$\rho_{G}^{\Gamma_{B}}\geq0$ if and only if $\rho_{G}$ is separable in
$\mathbb{C}_{A}^{2}\otimes\mathbb{C}_{B}^{q}$. Equivalently, the PPT-criterion
is necessary and sufficient to test separability in the described case (Wu
\cite{wu06}).

\item Wu \cite{wu06} considered generalized laplacians. Let $S$ be the set of
density matrices with nonnegative row sums and nonpositive off-diagonal
entries. If a density matrix $\rho\in S$ of dimension $n=pq$ is such that the
matrices $A_{ij}$ (as in Eq. \ref{1}) are line sum symmetric, then $\rho$ is
separable in $\mathbb{C}_{A}^{p}\otimes\mathbb{C}_{B}^{q}$. A matrix is
\emph{line sum symmetric} if the $i$-th column sum is equal to the $i$-th row
sum for each $i$. As a corollary, Wu \cite{wu06} proved that, if a density
matrix $\rho\in S$ of dimension $n=pq$ and with zero row sums is such that
$[\rho]_{i,j;i^{\prime},j^{\prime}}\neq0$ implies that $|i-i^{\prime}|\leq1$
then $\rho$ is separable in $\mathbb{C}_{A}^{p}\otimes\mathbb{C}_{B}^{q}$ if
and only if $\rho^{\Gamma_{B}}$ has zero row sums (Corollary 3). This result
generalizes separability of nearest point graphs. In fact, for a nearest point
graph also the condition $|j-j^{\prime}|\leq1$ is required. It is relevant to
notice that $\rho^{\Gamma_{B}}$ has zero row sum if and only if the
degree-criterion if satisfied.
\end{itemize}

\subsection{Equivalence of degree and PPT-criterion}

Here we prove that for laplacians the PPT-criterion is equivalent to the
degree criterion.

\bigskip

\noindent\textbf{Observation 1. }Let $\rho$ be a matrix acting on $\mathbb{C}%
_{A}^{p}\otimes\mathbb{C}_{B}^{q}$ and satisfying the PPT-criterion. Let
$x\otimes y$ be a separable vector in $\mathbb{C}_{A}^{p}\otimes\mathbb{C}%
_{B}^{q}$. Then the condition $\rho(x\otimes y)=0$ implies the condition
$\rho^{\Gamma_{B}}(x\otimes\overline{y})=0$. In fact, we have $\left( x\otimes
y\right) ^{\ast}\rho\left( x\otimes y\right) =\left( x\otimes\overline
{y}\right) ^{\ast}\rho^{\Gamma_{B}}\left( x\otimes\overline{y}\right) =0$, and
by the positivity of $\rho^{\Gamma_{B}}$ it follows that $\rho^{\Gamma_{B}}\left(
x\otimes\overline{y}\right) =0$.

\bigskip

Here the star denotes adjoint and the overbar denotes complex
conjugation.  A simple proof of Theorem 2 by Braunstein \emph{et
al.} \cite{braunstein06b} can be derived from this result, with $x$
and $y$ being equal to the all-ones vector.

\bigskip

\noindent\textbf{Observation 2. }As a consequence of Observation 1, if $\rho$
is a separable density matrix and
\[
\rho=\sum_{k=1}^{N}\left( x_{k}\otimes y_{k}\right) \left( x_{k}\otimes
y_{k}\right) ^{\ast}%
\]
is a separable decomposition of $\rho$ then, for any $k=1,2,...,N$, we have
the following conditions:

\begin{itemize}
\item $(x_{k}\otimes y_{k})\in range(\rho)$;

\item $(x_{k}\otimes\overline{y}_{k})\in range(\rho^{\Gamma_{B}})$.
\end{itemize}

\bigskip

\noindent\textbf{Theorem 1. }Let $\rho_{G}$ be the density matrix of a graph
$G$. Then $\rho_{G}$ satisfies the PPT-criterion if and only if it satisfies
the degree-criterion.

\bigskip

\begin{proof}We have $\rho_{G}\left( e\otimes e\right) =0$, because
$\rho_{G}$ is the laplacian of $G$ scaled by some coefficient, where
$e$ is the all-ones vector of the required dimension. Suppose that
the degree-criterion is satisfied. Then
$\rho_{G}^{\Gamma_{B}}=\rho_{G^{\Gamma _{B}}}$. Hence
$\rho_{G}^{\Gamma_{B}}$ is positive. It follows that $\rho_{G}$
satisfies the PPT-criterion. Suppose that the PPT-criterion is
satisfied. Then, by Observation 1, we have that
$\rho_{G}^{\Gamma_{B}}\left( e\otimes e\right) =0$. This is exactly
the degree-criterion on $\rho_{G}$.
\end{proof}

\subsection{Separability in $\mathbb{C}_{A}^{2}\otimes\mathbb{C}_{B}^{q}$}

Here we prove that the degree-criterion is necessary and sufficient to test
separability in $\mathbb{C}_{A}^{2}\otimes\mathbb{C}_{B}^{q}$ of density
matrices of graphs, therefore giving an alternative proof to a result of Wu
\cite{wu06}.

\bigskip

\noindent\textbf{Theorem 2. }Let $G$ be a graph on $n=2q$ vertices. Then
$\rho_{G}$ is separable in $\mathbb{C}_{A}^{2}\otimes\mathbb{C}_{B}^{q}$ if
and only if the degree-criterion is satisfied.

\bigskip

\begin{proof}The implication \textquotedblleft$\Rightarrow
$\textquotedblright\ is easily verified. We prove the implication
\textquotedblleft$\Leftarrow$\textquotedblright. If $G$ satisfies
the degree criterion then we can write
\[
\rho_{G}=L_{1}+L_{2}+L_{3},
\]
where%
\[
L_{1}:=\left(
\begin{array}
[c]{cc}%
X_{1} & 0\\
0 & 0
\end{array}
\right) ,L_{2}:=\left(
\begin{array}
[c]{cc}%
0 & 0\\
0 & X_{2}%
\end{array}
\right) \text{ and }L_{3}:=\left(
\begin{array}
[c]{cc}%
X_{3} & X_{4}\\
X_{4}^{T} & X_{3}%
\end{array}
\right)
\]
and $X_{1},...,X_{4}$ are appropriate matrices. Now, $L_{1}$ and
$L_{2}$ are trivially separable. The matrix $L_{3}$ is separable
because it is a PSD block-T\"{o}plitz matrix. Hence, $\rho_{G}$ is
separable.
\end{proof}.

\bigskip

\section{Concurrence}

In this section, we focus on the concurrence of density matrices of graphs.
The notion of concurrence was introduced by Hill and Wootters \cite{hill97}.
The concurrence of a density matrix acting on $\mathbb{C}_{A}^{p}%
\otimes\mathbb{C}_{B}^{q}$ is a quantity which is strictly larger than zero if
the state is entangled and zero if it is separable. Here is the definition. Let $|\psi\rangle_{AB}%
\in\mathbb{C}_{A}^{p}\otimes\mathbb{C}_{B}^{q}$. The \emph{concurrence} of
$|\psi\rangle_{AB}$ is denoted and defined as follows:%
\[
\mathcal{C(}\psi)=\sqrt{2(1-\text{tr}(\rho_{A}^{2}))},
\]
where%
\[
\rho_{A}=\text{tr}_{B}(|\psi\rangle_{AB}\langle\psi|).
\]
Let $\rho_{AB}$ be a density matrix acting on $\mathbb{C}_{A}^{p}%
\otimes\mathbb{C}_{B}^{q}$. The concurrence of $\rho_{AB}$ is denoted and
defined as
\[
\mathcal{C(}\rho_{AB})=\inf\left\{ \sum_{i}\omega_{i}\mathcal{C(}\psi
_{i}):\rho_{AB}=\sum_{i}\omega_{i}|\psi_{i}\rangle_{AB}\langle\psi_{i}%
|,0\leq\omega_{i}\leq1,\sum_{i}\omega_{i}=1\right\} .
\]
Let now $p = q = 2$ and
\[
\sigma_{y}=-i|1\rangle\langle2|+i|2\rangle\langle1|,
\]
where $|1\rangle$ and $|2\rangle$ are the eigenvectors of the matrix
\[
\sigma_{z}=\left(
\begin{array}
[c]{rr}%
1 & 0\\
0 & -1
\end{array}
\right) ,
\]
corresponding to the eigenvalues $1$ and $-1$, respectively. An analytical formula for
$\mathcal{C(}\rho_{AB})$, is given by
\[
\mathcal{C(}\rho_{AB})=\max\{0,\lambda_{1}-\lambda_{2}-\lambda_{3}-\lambda
_{4}\},
\]
where $\lambda_{1},\lambda_{2},\lambda_{3}$ and $\lambda_{4}$ are the square
roots of the eigenvalues of $\rho_{AB}\widetilde{\rho}_{AB}$ arranged in
decreasing order and
\[
\widetilde{\rho}_{AB}:=(\sigma_{y}\otimes\sigma_{y})\rho_{AB}^T%
(\sigma_{y}\otimes\sigma_{y}).
\]
The importance of the concurrence stems from its relation with the so-called
\emph{entanglement of formation}, the most widely accepted measure of
entanglement (Bennett \emph{et al.} \cite{bennett96}; see also Plenio and
Virmani \cite{plenio06}). For a pure state (that is a state of the form
$P\left[ \left\vert \psi\right\rangle \right] $) of a system $\emph{S}_{AB}%
$, a good measure of entanglement is the entropy of the density matrix
associated with one of the two subsystems. Choosing the system $\emph{S}_{A}$,
this can be written as%
\[
E(\psi):=-\text{tr}(\rho_{A}\log_{2}\rho_{A}),
\]
where%
\[
\rho_{A}=\text{tr}_{B}(P\left[ \left\vert \psi\right\rangle \right] ).
\]
For a mixed state $\rho$, the entanglement of formation is defined by%
\[
E_{f}(\rho):=\min\sum_{i}p_{i}E(\psi_{i}),
\]
where the minimum is taken over all pure states decompositions of the density
matrix $\rho$. It is evident that computing $E_{f}$ is in general not an easy
task. Explicit formulas are only known for very specific classes of states. An
example consists of Werner states (Vollbrecht and Werner \cite{voll01}). The
role of concurrence is explained by the following result (Wootters
\cite{wootters98}). Let $\rho$ be a mixed density matrix of dimension $4$.
Then%
\[
E_{f}(\rho)=H\left( \frac{1}{2}+\frac{1}{2}\sqrt{1-\mathcal{C(}\rho)^{2}%
}\right) ,
\]
where
\[
H(x)=-x\ln x-(1-x)\ln(1-x),
\]
is the standard information-theoretic entropy. Remarkably, $E_{f}(\rho)$
increases monotonically as a function of $\mathcal{C(}\rho)$.

For density matrices of graphs of dimension $4$ the situation can be
described as follows. There are twelve nonisomorphic graphs on $4$
vertices. Seven of these graphs can have entangled
density matrices. The tables below present
these graphs and their respective concurrence:%

\[%
\begin{tabular}
[c]{c}%
$%
\begin{tabular}
[c]{ccccccc}%
$1/3$ & & $1/3$ & & $1/5$ & & $1$\\%
{\includegraphics[
height=0.4465in,
width=0.3609in
]
{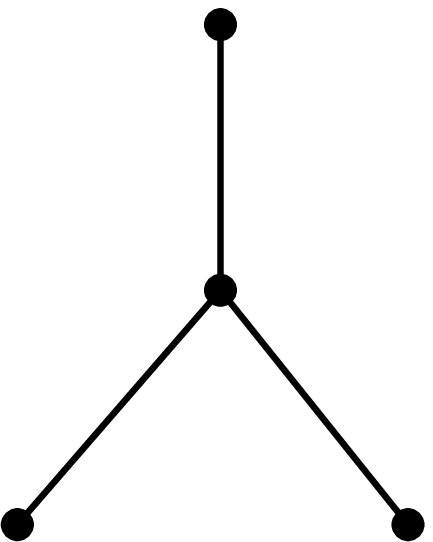}
}%
& &
{\includegraphics[
height=0.2626in,
width=0.3651in
]
{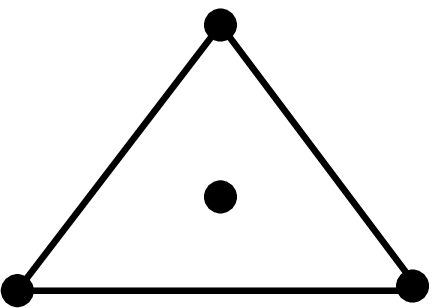}
}%
& &
{\includegraphics[
height=0.5202in,
width=0.3728in
]%
{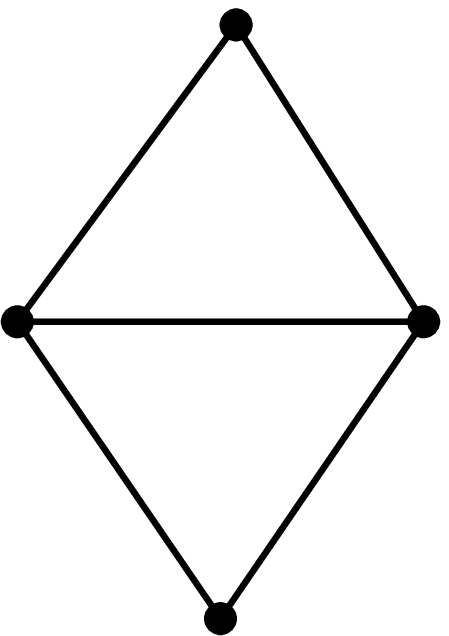}
}%
& &
{\includegraphics[
height=0.227in,
width=0.3516in
]
{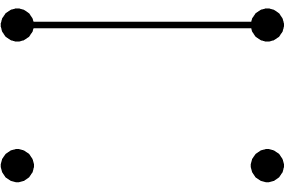}
}%
\end{tabular}
$\\
\\
$%
\begin{tabular}
[c]{ccccc}%
$1/4$ & & $1/2$ & & $1/3$\\%
{\includegraphics[
height=0.46in,
width=0.3228in
]%
{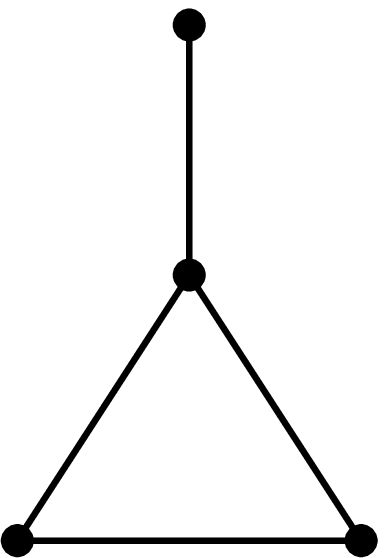}%
}%
& &
{\includegraphics[
height=0.2728in,
width=0.4109in
]%
{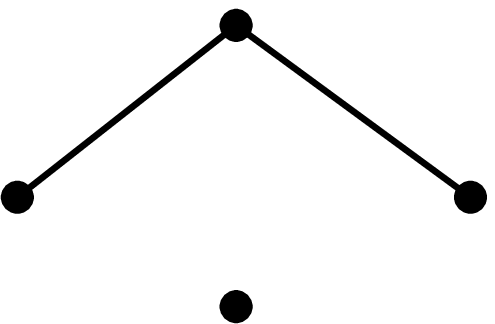}%
}%
& &
{\includegraphics[
height=0.2236in,
width=0.4592in
]%
{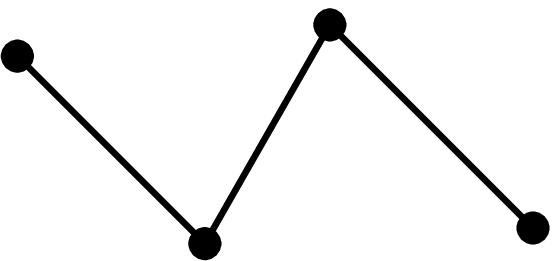}%
}%
\end{tabular}
$%
\end{tabular}
\]
Notice that in all cases the value of the concurrence is 1 over the number of edges.
Moreover, one easily sees that the optimal decomposition of the density matrices
of these graphs just corresponds to the decomposition of the combinatorial
laplacians as sums of laplacians of 1-edge graphs.

This motivates the following definitions.

{\definition Let $G = (V,E)$ be a graph, with its $n = pq$ vertices arranged in a $p \times q$ array.
We call an egde $e \in E$ \emph{separable} if the density matrix of the 1-edge graph
$G_e = (V,\{e\})$ is separable. We call $e \in E$ \emph{matched} if $e \in E^{\prime}$
and \emph{unmatched} otherwise. }

\bigskip

Recall that $E^{\prime}$ is the set of edges of the partially transposed graph $G^{\prime}$.
Thus an edge is matched if and only if it is part of a criss-cross\footnote{
A \emph{criss-cross} is a set
$\{\{(k,i),(l,j)\},\{(k,j),(l,i)\}\}$ of two edges belonging to
a set of (vertex-disjoint) entangled edges
on $n=pq$ vertices
 (see also \cite{braunstein06a}).}
 or it is separable.
Since graphs consisting of a criss-cross give rise to separable density matrices,
we have the following results.

\bigskip

\noindent\textbf{Observation 3. }Let $G = (V,E)$ be as above.
Let further $E_1 \subset E$ be the subset of all matched edges. Then the graph
$G_1 = (V,E_1)$ has a separable density matrix.

\bigskip

\noindent\textbf{Observation 4. }Let $G = (V,\{e\})$ be a 1-edge graph, with its $n = pq$ vertices arranged
in a $p \times q$ array, and let $\rho_e$ be its density matrix. Then the concurrence of $\rho_e$ is given by
0 if $e$ is a separable egde and by 1 if $e$ is not separable, and hence unmatched.

{\corollary Let $G = (V,E)$ be as above. Let $n_1$ be the number of matched egdes of $G$
and $n_2$ be the number of unmatched edges. Then the concurrence of the density matrix
$\rho$ of $G$ is bounded from above by
\[ \mathcal{C(}\rho) \leq \frac{n_2}{n_1+n_2}.
\]
In particular, for any graph $G$ with density matrix $\rho$ we have $\mathcal{C(}\rho) \leq 1$.
}

\bigskip

\begin{proof} Assume the above notations. Let $\rho_1,\rho_2$ be the
density matrices of the graphs $(V,E_1)$ and $(V,E_2)$,
respectively, where $E_1$ is the set of matched edges and $E_2$ the
set of unmatched edges. Then the density matrix $\rho$ of $G$ is
given by the convex combination $\rho = \frac{n_1}{n_1+n_2}\rho_1 +
\frac{n_2}{n_1+n_2}\rho_2$ and the density matrix $\rho_2$ is given
by $\rho_2 = \frac{1}{n_2} \sum_{e \in E_2} \rho_e$, where $\rho_e$
is the density matrix of the 1-edge graph $(V,\{e\})$. By convexity
of the concurrence and by Observations 3 and 4 we obtain
$$\mathcal{C(}\rho) \leq \frac{n_1}{n_1+n_2}\mathcal{C(}\rho_1) + \sum_{e \in E_2} \frac{1}{n_1+n_2}\mathcal{C(}\rho_e)
= \frac{n_2}{n_1+n_2}.$$
\end{proof}

For all of the above graphs on four vertices either the density matrix is separable or we have $n_2 = 1$,
in which case 1 over the number of edges is an upper bound for the concurrence. As can be seen from the
table, the bound is actually achieved.

The concurrences of graph density matrices which have rank 2 are listed in \cite{Hildebrand0612064}.
Examples IVb,IVc and IX in \cite{Hildebrand0612064} or the tally-mark\footnote{
 A  \emph{tally-mark} is a set

\bigskip

$\{(k,i_{1}),(l,i_{2})\},\{(k,i_{2}),(l,i_{3})\},...,\{(k,i_{s+1}%
),(l,i_{s+2})\},\{(k,i_{s+2}),(l,i_{1})\}$

\bigskip

of $s+2$ edges, where $k<l$, $s\geq0$ and $i_{1}<i_{2}<\cdots<i_{s+2}$,
belonging to a set of (vertex-disjoint) entangled edges
 on $n=pq$ vertices (see also \cite{braunstein06a}).
 Note that a criss-cross is a tally-mark with two edges.}
  show that
in general the upper bound is not exact, even for graphs on $2 \times 3$ arrays.

{\definition We call a graph $G = (V,E)$ \emph{maximally entangled} if the concurrence of its density matrix
$\rho$ is given by $\mathcal{C(}\rho) = 1$. }

All edges of a maximally entangled graph must hence be unmatched.

Some natural questions arise:

\begin{itemize}
\item Are there nonisomorphic graphs with the same concurrence?

\item How can graphs with rational concurrence be characterized?

\item Is the concurrence of $\rho_{G}$ related to specific combinatorial
properties of $G$?

\item Can the set of edges of a graph $G= (V,E)$ with density matrix $\rho$
always be partitioned in two subsets $E_1,E_2$ such that the density matrix
of $(V,E_1)$ is separable, $(V,E_2)$ is maximally entangled and
$\mathcal{C(}\rho) = \frac{\#E_2}{\#E}$?

\item How can maximally entangled graphs be characterized?

\item Does there exist a class of density matrices of graphs for which testing
separability is a difficult problem? Maybe, the existence of such a
class would provide a transparent proof that detecting entanglement
is hard.

\end{itemize}

Unfortunately, explicit formulae for computing the concurrence of density
matrices are known so far only for dimensions $n \leq 4$ (see, Rungta \emph{et al.}
\cite{rungta01}, Li and Zhu \cite{li03}, and Mintert \emph{et al.}
\cite{mintert04}) and for density matrices of rank 2 \cite{Hildebrand0612064}.
This is an obstacle when thinking about the questions
above. Nevertheless, one can still hope to find an \emph{ad-hoc} formula for
$\mathcal{C(}\rho)$, when $\rho$ is the density matrix of a graph. In fact, it
may well be that the optimal decompositions of $\rho_{G}$ in pure states are very
special. Finding such a formula would be interesting in view of potential
generalizations.

Putting the concurrence on a side, one may ask if there is some entanglement
measure specifically tailored for $\rho_{G}$. Considering the apparent success
of the degree-criterion, a naive\emph{ }measure would be the normalized
Euclidean norm $EN(\rho_{G}):=\left\Vert \Delta(G)-\Delta(G^{\Gamma_{B}%
})\right\Vert $. The \emph{logarithmic negativity} is a well-known entanglement
measure and it is defined by $LN(\rho_{G}):=\log_{2}(1+2\mathcal{N}(\rho
_{G}))$, where $\mathcal{N}(\rho_{G})$ is the sum of the magnitudes of all
negative eigenvalues of $\rho_{G}^{\Gamma_{B}}$ (see Vidal and Werner
\cite{vidal02}). There are examples of graphs $G$ and $H$ for which
$EN(\rho_{G})=EN(\rho_{H})$ but $LN(\rho_{G})\neq LN(\rho_{H})$ (Ghosh
\cite{ghosh}).

\section{Conclusion}

We have proven that the degree-criterion is equivalent to the
PPT-criterion. It is thus in general not sufficient for separability of density
matrices of graphs. As a matter of fact, we have provided a counterexample within
graphs having isolated vertices. Nevertheless, we have been able to prove the sufficiency of
the degree-criterion when one of the subsystems has dimension two.
In particular, as a corollary of Theorem 2, one can easily obtain the separability of criss-crosses and
tally-marks.

We have also considered the concurrence as a possible entanglement measure of density
matrices of graphs. There could be more suitable entanglement measures for such kind of
states, especially because no explicit formula for concurrence is known when
$n>4$ and the rank of the density matrix is exceeding 2. Further studies are in order on the subject of density matrices of
graphs. However we believe that such states provide a restricted testing
ground for better understanding techniques and concepts employed in more
general settings.




\begin{thebibliography}{99} %


\bibitem {alber01}G. Alber \emph{et al.} (Eds.), \emph{Quantum Information: An
Introduction to Basic Theoretical Concepts and Experiments,} Series: Springer
Tracts in Modern Physics, Vol. 173, 2001.

\bibitem {bennett96}C. H. Bennett, D. P. DiVincenzo, J. A. Smolin, and W. K.
Wootters, Mixed-state entanglement and quantum error correction, \emph{Phys.
Rev. A} (3) \textbf{54} (1996), no. 5, 3824--3851, quant-ph/9604024.

\bibitem {braunstein06a}S. L. Braunstein, S. Ghosh and S. Severini, The
laplacian of a graph as a density matrix: a basic combinatorial approach to
separability of mixed states, to appear in \emph{Ann. Comb.}, quant-ph/0406165.

\bibitem {braunstein06b}S. L. Braunstein, S. Ghosh, T. Mansour, S. Severini,
and R. C. Wilson, Some families of density matrices for which separability is
easily tested, \emph{Phys. Rev. A,} \textbf{73}:1, 012320 (2006), quant-ph/0508020.

\bibitem {brus02}D. Bru\ss , Characterizing entanglement. Quantum information
theory. \emph{J. Math. Phys. }\textbf{43} (2002), no. 9, 4237--4251. quant-ph/0110078.

\bibitem {chen02}K. Chen and L.-A. Wu, A matrix realignment method for
recognizing entanglement, \emph{Quantum Inf. Comput.} \textbf{3} (2003), no.
3, 193--202, quant-ph/0205017.

\bibitem {duan}L. M. Duan, G. Giedke, J. I. Cirac, and P. Zoller,
Inseparability criterion for continuous variable systems, \emph{Phys. Rev.
Lett.} \textbf{84}, 2722 (2000), quant-ph/9908056.

\bibitem {ghosh}S. Ghosh, Personal communication, June 2006.

\bibitem {godsil01}C. Godsil and G. Royle, Algebraic graph theory. Graduate
Texts in Mathematics, 207. \emph{Springer-Verlag, New York,} 2001.

\bibitem {gro88}M. Gr\"{o}tschel, L. Lov\'{a}sz, and A. Schrijver,
\emph{Geometric Algorithms and Combinatorial Optimization} (Springer-Verlag,
Berlin, 1988).

\bibitem {gurvits03}L. Gurvits, Classical deterministic complexity of Edmonds'
Problem and quantum entanglement, \emph{Proceeding of the thirty-fifth ACM
Symposium on Theory of Computing (ACM Press, New York),} 2003, 10-19, quant-ph/0303055.

\bibitem {Hildebrand0612064}R. Hildebrand, Concurrence of Lorentz-positive maps, quant-ph/0612064.

\bibitem {hill97}S. Hill and W. K. Wootters, Entanglement of a Pair of Quantum
Bits, \emph{Phys. Rev. Lett.} \textbf{78}, 5022 (1997).

\bibitem {ho96}M. Horodecki, P. Horodecki, and R. Horodecki, Separability of
mixed states: necessary and sufficient conditions, \emph{Phys. Lett. A}
\textbf{223} (1996), no. 1-2, 1--8, quant-ph/9605038.

\bibitem {ho00}P. Horodecki and M. Lewenstein, Bound entanglement and
continuous variables, \emph{Phys. Rev. Lett.} \textbf{85} (2000), no. 13,
2657--2660, quant-ph/0001035.

\bibitem {ioannou06}L. M. Ioannou, Deterministic Computational Complexity of
the Quantum Separability Problem, quant-ph/0603199.

\bibitem {li03}Y.-Q. Li, G.-Q. Zhu, Concurrence Vectors for Entanglement of
High-dimensional Systems, quant-ph/0308139.

\bibitem {mintert04}F. Mintert, M. Ku\'{s}, and A. Buchleitner, Concurrence of
mixed multipartite quantum states, \emph{Phys. Rev. Lett.} \textbf{95} (2005),
no. 26, 260502,quant-ph/0411127.

\bibitem {mohar88}B. Mohar, \emph{The Laplacian spectrum of graphs.} Graph
theory, combinatorics, and applications. Vol. 2 (Kalamazoo, MI, 1988),
871---898, Wiley-Intersci. Publ., Wiley, New York, 1991.

\bibitem {nielsen00}M. Nielsen and I. Chuang, \emph{Quantum Computation and
Quantum Information} (Cambridge University Press, Cambridge, 2000).

\bibitem {peres96}A. Peres, Separability criterion for density matrices,
\emph{Phys. Rev. Lett.} \textbf{77} (1996), no. 8, 1413--1415, quant-ph/9604005.

\bibitem {plenio06}M. B. Plenio and S. Virmani, An introduction to
entanglement measures, quant-ph/0504163.

\bibitem {rudolph02}O. Rudolph, Further results on the cross norm criterion
for separability, \emph{Quantum Inf. Process.} \textbf{4} (2005), no. 3,
219--239, quant-ph/0202121.

\bibitem {rungta01}P. Rungta, V. Bu\v{z}ek, Carton M. Caves, M. Hillery, and
G. J. Milburn, Universal state inversion and concurrence in arbitrary
dimensions, Phys. Rev. A (3) 64 (2001), no. 4, 042315, quant-ph/0102040.

\bibitem {simon00}R. Simon, Peres-Horodecki Separability Criterion for
Continuous Variable Systems, \emph{Phys. Rev. Lett.} \textbf{84}, 2726--2729
(2000), quant-ph/9909044.

\bibitem {mancini06}S. Mancini and S. Severini, The Quantum Separability
Problem for Gaussian States, \emph{LMCS06 (Logic, models and computer
science), Camerino, Italy,} to appear in \emph{ENTCS}, cs.CC/0603047.

\bibitem {vidal02}G. Vidal and R. Werner, Computable measure of entanglement,
\emph{Phys. Rev. A} \textbf{65}, 032314 (2002).

\bibitem {voll01}K. G. H. Vollbrecht and R. F. Werner, Entanglement measures
under symmetry, \emph{Phys. Rev. A} \textbf{64}, 062307 (2001), quant-ph/0010095.

\bibitem {wu06}C. W. Wu, Conditions for separability in generalized Laplacian
matrices and diagonally dominant matrices as density matrices, \emph{Phys.
Lett. A }\textbf{351} (2006), no. 1-2, 18--22, quant-ph/0508163.

\bibitem {wootters98}W. K. Wootters, Entanglement of Formation of an Arbitrary
State of Two Qubits, \emph{Phys. Rev. Lett.} \textbf{80}, 2245 (1998).
\end{thebibliography}
\end{document}